# Dielectric constants of Ir, Ru, Pt, and IrO$_2$: Contributions from bound charges

W. S. Choi, S. S. A. Seo, K. W. Kim, and T. W. Noh[*]

*ReCOE & FPRD, Department Physics and Astronomy, Seoul National University, Seoul 151-747, Korea*

M. Y. Kim

*FPRD, Department of Physics and Astronomy, Seoul National University, Seoul 151-747, Korea*

S. Shin

*Samsung Advanced Institute of Technology, Suwon 440-600, Korea*

We investigated the dielectric functions $\hat{\varepsilon}(\omega)$ of Ir, Ru, Pt, and IrO$_2$, which are commonly used as electrodes in ferroelectric thin film applications. In particular, we investigated the contributions from bound charges $\hat{\varepsilon}^b(\omega)$, since these are important scientifically as well as technologically: the $\varepsilon_1^b(0)$ of a metal electrode is one of the major factors determining the depolarization field inside a ferroelectric capacitor. To obtain $\varepsilon_1^b(0)$, we measured reflectivity spectra of sputtered Pt, Ir, Ru, and IrO$_2$ films in a wide photon energy range between 3.7 meV and 20 eV. We used a Kramers–Kronig transformation to obtain real and imaginary dielectric functions, and then used Drude–Lorentz oscillator fittings to extract $\varepsilon_1^b(0)$ values. Ir, Ru, Pt, and IrO$_2$ produced experimental $\varepsilon_1^b(0)$ values of 48 ± 10, 82 ± 10, 58 ± 10, and 29 ± 5, respectively, which are in good agreement with values obtained using first-principles calculations. These values are much higher than those for noble metals such as Cu, Ag, and Au because transition metals and IrO$_2$ have such strong

**$d$–$d$ transitions below 2.0 eV. High $\varepsilon_1^b(0)$ values will reduce the depolarization field in ferroelectric capacitors, making these materials good candidates for use as electrodes in ferroelectric applications.**

## I. INTRODUCTION

Recently, studies have focused on the physics at the interface of a ferroelectric (FE) material and a metal electrode,[1,2] in particular on the depolarization field and its effects.[3,4] In a FE capacitor, polarization charges are induced at the metal/FE interface. These charges are screened by free carriers inside the metal. However, screening is not complete due to the finite nature of charge screening, resulting in some effective charges at the interface. These charges induce an electric field pointing in a direction opposite to the polarization inside the capacitor. This depolarization field, $E_d$, is a very important physical quantity and governs the critical thickness[3,4] and domain dynamics[5] of ultrathin FE films. When FE film thickness is reduced, the field's effects become more substantial.

Quantitative determination of $E_d$ values requires knowing the values of some physical quantities of the metal electrode as well as of the FE layer. Mehta et al.[6] used electrostatic calculations to express the depolarization field as

$$E_d = -\frac{P}{\varepsilon_0 \varepsilon_F} \left( \frac{2\varepsilon_F/d}{2\varepsilon_F/d + \varepsilon_e/\lambda} \right), \qquad (1)$$

where $P$ is the polarization inside the FE layer, $d$ is thickness, and $\varepsilon_F$ is the $dc$ dielectric constant of the FE layer. Physical properties of the metal electrode layer are summarized in the $\varepsilon_e/\lambda$ term, where $\lambda$ is the screening length and $\varepsilon_e$ is the *dc* dielectric constant due to bound charges, $\varepsilon_1^b(\omega = 0)$.[7] When an external electric field is applied to the capacitor, free charge carriers accumulate at the interface. In addition, bound charges inside the metal electrode shift position, so $\varepsilon_1^b(0)$ affects the effective screening at the interface. Therefore, understanding the intriguing physical phenomena occurring within ultrathin FE films requires accurate knowledge of the metal electrode's $\varepsilon_1^b(0)$ value.

Most commercial applications of FE devices use transition metals (such as Pt, Ir, and Ru) or transition metal oxides (such as $IrO_2$ and $SrRuO_3$) as metal electrodes.[8,9] Theoretically, estimation of $\varepsilon_1^b(0)$ values can be straightforward: we can calculate interband transitions and sum their contributions to the *dc* dielectric constant. However, most theoretical calculations for metals have not estimated these interband transition contributions separately. Experimental determination of $\varepsilon_1^b(0)$ is not trivial for any metal, since the free carrier Drude response is much greater and screens $\varepsilon_1^b(0)$ behavior. To the best of our knowledge, no report has provided $\varepsilon_1^b(0)$ values for these transition metals or $IrO_2$.

Ehrenreich and Philipp[10] used optical spectroscopy to measure $\varepsilon_1^b(0)$ values for Ag and Cu; they experimentally measured complex dielectric functions $\hat{\varepsilon}(\omega)$ ($\equiv \varepsilon_1(\omega) + i\varepsilon_2(\omega)$, and $\varepsilon_2(\omega) = 4\pi\sigma_1(\omega)/\omega$) of the noble metals and fit them with the Drude free carrier response and Lorentz oscillators. They were able to separate $\varepsilon_1^b(0)$ values by subtracting the Drude free carrier response from $\varepsilon_1(\omega)$ and taking the limit of $\omega \to 0$. We used the same procedure to experimentally determine the $\varepsilon_1^b(0)$ value of $SrRuO_3$ and found that $\varepsilon_1^b(0) \approx 8.17$.[4] Although $SrRuO_3$ had a lower carrier density than the noble metals, it had a higher $\varepsilon_1^b(0)$ value. Therefore, determination of accurate $\varepsilon_1^b(0)$ values for other metal electrodes (such as Ir, Ru, Pt, and $IrO_2$) is essential in many theoretical investigations of ultrathin FE capacitor physics, as well as actual device applications. This paper reports $\varepsilon_1^b(0)$ values for Ir, Ru, Pt, and $IrO_2$, which we obtained using optical spectroscopy and first-principles calculations. We observed optical transitions in their optical spectra at frequencies under 2.0 eV. These low-energy optical transitions are mainly caused by optical processes involving $d$-electrons. Since the low-energy optical transitions greatly overlapped with free carrier responses, it was not easy to separate $\varepsilon_1^b(0)$ from free carrier contribution to $\varepsilon_1(0)$.[11,12] We attempted to obtain accurate $\hat{\varepsilon}(\omega)$ values by measuring reflectivity spectra in a wide photon energy region (from 3.7 meV to 20 eV) and by conducting a Kramers–Kronig (K–K) analysis. We also conducted independent ellipsometry measurements to verify consistency. We used Drude–Lorentz oscillator fittings to clearly separate free carrier responses and optical transitions, and evaluated their contributions to $\varepsilon_1(0)$. We also performed first-principles calculations on the dielectric responses of the transition metals and $IrO_2$; results were in good agreement with experimentally determined $\varepsilon_1^b(0)$ values.

Section II of this paper describes our experimental and theoretical methods. Section III presents our experimental and theoretical results. It also explains how we analyzed our experimental optical spectra, and compares the experimental and theoretical results of $\hat{\varepsilon}(\omega)$ for each material, showing that the commonly used electrode materials have quite high $\varepsilon_1^b(0)$ values. Section IV briefly discusses the origin of these high $\varepsilon_1^b(0)$ values. Section V summarizes the study's main conclusions.

## II. EXPERIMENTAL & THEORETICAL METHODS

## A. Film fabrication

We used a sputtering technique to deposit Ir, Ru, and Pt films on Si substrates and to deposit $IrO_2$ films on $Al_2O_3$ (0001) substrates. Si substrate surfaces had thin natural $SiO_2$ amorphous layers, on which we grew the polycrystalline transition metals. Since Pt and Ir have cubic crystal symmetries, they should have an isotropic $\hat{\varepsilon}(\omega)$, meaning that optical measurements of these polycrystalline films can provide the same information as that obtained from the single crystalline samples. In contrast, Ru has a hexagonal close-packed crystal structure, and $IrO_2$ has a rutile structure, which has a hexagonal net in the *ab* plane. We used $Al_2O_3$ (0001) substrates to grow $IrO_2$ films, since the hexagonal symmetry of their surfaces would produce better $IrO_2$ films. Instead of growing epitaxial Ru and $IrO_2$ films, we chose to grow polycrystalline films because they are more commonly used in actual applications. In this case, the obtained optical spectra should be interpreted as an effective dielectric function $\hat{\varepsilon}(\omega)$ of randomly oriented $IrO_2$ grains. Film thickness was measured using cross-sectional SEM, and ranged from 366 to 415 nm. [Results indicated that film skin depth should be much smaller than film thicknesses in most of the measured photon energy region.]

## B. Optical measurements

We obtained reflectance spectra of electrode materials in a wide photon energy range (3.7 meV to 20 eV) at room temperature. We used two different Fourier transform spectrophotometers to obtain spectra between 3.7 meV and 1.5 eV, a grating monochromator to obtain spectra between 0.6 eV and 6.9 eV, and synchrotron radiation at the Pohang Light Source (PLS) to obtain spectra above 4.8 eV. In addition to these reflectance measurements, we used spectroscopic ellipsometry to independently measure optical functions of films from 1.5 eV to 5.0 eV.

Figure 1 shows the reflectance spectra of Ir, Ru, and Pt films from 3.7 meV to 20 eV. All measured reflectance spectra exhibited typical metallic behavior in the low photon energy region: almost 100% reflectance. The low energy high reflectance in noble metals is close to 100% in the visible region and begins to decrease near plasma frequencies in the ultraviolet region. However, in this study we observed some dips in reflectance spectra in or below the visible region of the transition metals and $IrO_2$, probably originating from electronic transitions. Section III addresses these electronic transitions in detail with corresponding conductivity spectra.

### C. Kramers–Kronig analysis

The native oxide layer that possibly formed on transition metal film surfaces might have affected reflectance spectra.[13] To test for such an effect, we used results from our spectroscopic ellipsometry measurements to analyze thin film geometry. Data indicated that native oxide layers (e.g., $IrO_2$, $RuO_2$) that formed on transition metal films were likely to have thicknesses of less than 1 nm. Such thin native oxide layers should produce only negligible changes in reflectance spectra and K–K analyses, and should fall within our experimental errors.

We conducted K–K analyses to obtain $\hat{\varepsilon}(\omega)$ from measured reflectance spectra. Since skin depth was much smaller than film thickness in most photon energy regions, we could ignore reflected light coming from the film/substrate. For K–K analyses, we extrapolated reflectance spectra in low frequency regions below our measurements using the Hagen–Rubens relationship.[14] For frequency regions above 20 eV, we extended reflectance at 20 eV to 30 eV and then assumed $\omega^{-4}$ dependence. Optical functions independently obtained using spectroscopic ellipsometry were quite consistent with those obtained using K–K analyses, indicating the validity of the extrapolations we used in K–K analyses.

### D. First-principles calculations

We performed accurate first-principles calculations employing a full-potential linearized augmented plane–wave (FLAPW) method[15] within local density approximation. We treated core states fully relativistically, and for valence states we treated the spin orbit coupling interaction in a self-consistent second variational way. We employed an energy cutoff of 13.0 Rydberg for the plane–wave basis and used $l \leq 8$ spherical harmonics inside the muffin-tin radius. We adopted crystal structures and lattice constants from experiments for the simple metals. $IrO_2$ has six atoms in a tetragonal unit cell: two metal atoms at (0, 0, 0) and (0.5, 0.5, 0.5) and four oxygen atoms at ±(u, u, 0) and ±(0.5 + u, 0.5 − u, 0.5). We obtained the optimized atomic position of oxygen of u = 0.309 for $IrO_2$ using atomic force calculations assuming the experimental lattice constants of a = 8.500 a.u. and c/a = 0.701.

We calculated dielectric functions within an electric dipole approximation. Imaginary parts of the dielectric functions for interband transitions are expressed as[16]

$$\varepsilon_2^{interband}(\omega) = \frac{8\pi^2 e^2}{m^2 \omega^2 V} \sum_{c,v} \sum_{\mathbf{k}} |\langle c,\mathbf{k}|\hat{\mathbf{e}} \cdot \mathbf{p}|v,\mathbf{k}\rangle|^2 \times \delta[E_c(\mathbf{k}) - E_v(\mathbf{k}) - \hbar\omega], \quad (2)$$

where $c$ and $v$ are the conduction and valence states, respectively, |$n$, **k**> are the FLAPW eigenstates, **p** is the momentum operator, and $\hat{\mathbf{e}}$ is the external field vector. Integration in **k** space used a modified tetrahedron method with a grid containing about 500 and 3,000 **k**-points in IBZ for FLAPW and optical properties calculations, respectively. We used K–K transformations of the calculated $\varepsilon_2^{interband}(\omega)$ to obtain $\varepsilon_1^{interband}(\omega)$.[17] Our

calculations did not include Drude-type intraband transitions (which are often added to provide better descriptions of low-energy optical properties, especially for metals) so that results could be directly compared with experimental results of the contribution of bound electrons.

### III. RESULTS AND ANALYSIS

### A. Experimental $\hat{\varepsilon}^b(\omega)$

We analyzed the experimentally measured $\hat{\varepsilon}(\omega)$ using Drude–Lorentz oscillator fittings. Figure 2 displays results for the Ir film. [Figures 3, 4, and 5 provide similar results for Ru, Pt, and IrO$_2$ films, respectively.] The empty circles in Fig. 2(a) show $\sigma_1(\omega)$ obtained from K–K analysis. We fitted these $\sigma_1(\omega)$ with a Drude term and a summation of the Lorentz oscillators:

$$\sigma_1(\omega) = \frac{e^2}{m^*} \frac{n_D \gamma_D}{\omega^2 + \gamma_D^2} + \frac{e^2}{m^*} \sum_j \frac{n_j \gamma_j \omega^2}{\left(\omega_j^2 - \omega^2\right)^2 + \gamma_j^2 \omega^2}, \qquad (3)$$

where $e$ and $m^*$ are the electron charge and effective electron mass, respectively. For the Drude term, $n_D$ and $\gamma_D$ are the density and scattering rate of free electrons, respectively. For the Lorentz oscillators, $n_j$, $\gamma_j$, and $\omega_j$ are the electron density, scattering rate, and resonant frequency of the $j$-th oscillator, respectively. Table I presents a summary of $\gamma_j$, $\omega_j$, and plasma frequencies ($\omega_{pj}^2 = 4\pi n_j e^2/m^*$) values of the Lorentz oscillator terms for Ir, Ru, Pt, and IrO$_2$.

In Fig. 2(a), the thin solid lines in the lower part display the contribution of each Lorentz oscillator to $\sigma_1(\omega)$. The thick dashed line provides a summation of all Lorentz oscillator contributions, and the dotted line displays the contribution from the Drude term. The excellent agreement between the fitting curve (*i.e.*, the thick solid line) and the experimental data allowed us to verify the validity of this analysis. Using this result, we separated the dielectric constant from bound charge contribution $\hat{\varepsilon}^b(\omega)$:

$$\hat{\varepsilon}^b(\omega) = 1 + \frac{4\pi e^2}{m^*} \sum_j \frac{n_j}{\left(\omega_j^2 - \omega^2\right)^2 - i\gamma_j \omega}. \quad (4)$$

Figure 2(b) presents real and imaginary parts of $\hat{\varepsilon}^b(\omega)$ for Ir thin films.

## B. Comparison between $\hat{\varepsilon}^b(\omega)$ and $\hat{\varepsilon}^{interband}(\omega)$

Under electric dipole approximation, each allowed interband transition can contribute to $\hat{\varepsilon}(\omega)$, as shown in Eq. (2). These transitions can be approximately rewritten in Lorentz oscillator terms, as shown in Eq. (3). Therefore, the theoretical $\hat{\varepsilon}^{interband}(\omega)$, obtained by taking into account interband transitions, should correspond to dielectric function due to the bound charge, *i.e.* $\hat{\varepsilon}^b(\omega)$.[7] Calculations of $\varepsilon_1^{interband}(\omega)$ may contain an overshooting problem, which might be closely related to the small gap in $\varepsilon_2^{interband}(\omega)$; the origin of these features is not yet understood. The theoretical $\hat{\varepsilon}^{interband}(\omega)$ have relatively clear and sharp peaks compared to the experimental $\hat{\varepsilon}^b(\omega)$. One of the reasons might be due to the facts that our polycrystalline films might have numerous defects (such as vacancies, dislocations, and grain boundaries) and related disorders.

According to the Drude model, the free carrier causes the dielectric constant to decrease to a high negative value as photon energy approaches zero. When we exclude this free carrier contribution and set a limit of $\omega \rightarrow 0$, $\varepsilon_1^b(\omega)$ produces a finite positive number, which is the dielectric constant of bound charge, $\varepsilon_1^b(0)$. We can also use the same limit to theoretically determine the dielectric constant due to contributions of interband transitions, $\varepsilon_1^{interband}(0)$. Table II summarizes the experimental and calculated results of the dielectric constants of bound charges; the theoretical $\varepsilon_1^{interband}(0)$ values are in good agreement with the experimental $\varepsilon_1^b(0)$. The following subsections provide a detailed comparison of $\widehat{\varepsilon}(\omega)$ and $\widehat{\varepsilon}^{interband}(\omega)$ for the metal electrode materials commonly used in FE applications.

### *1. Ir film*

Figure 2 presents the optical spectra of Ir. According to previous optical studies,[18] Ir has absorption peaks around 1.05, 1.9, 3.1, and 4.12 eV. Our $\sigma_1(\omega)$ spectra, shown in Fig. 2(a), also have peaks at similar positions. In the previous work, the 1.0 eV peak could be seen weakly in $\sigma_1(\omega)$, but not in $\varepsilon_2(\omega)$. However, by carefully excluding the response of the free carriers, our data clearly indicate the peak at 1.0 eV in both $\sigma_1(\omega)$ and $\varepsilon_2^b(\omega)$.

The theoretically obtained $\widehat{\varepsilon}^{interband}(\omega)$ spectra shown in Fig. 2(c) are in very good agreement with $\widehat{\varepsilon}^b(\omega)$. First, the overall shapes of both $\varepsilon_1^{interband}(\omega)$ and $\varepsilon_2^{interband}(\omega)$ spectra agree with those of $\varepsilon_1^b(\omega)$ and $\varepsilon_2^b(\omega)$, respectively. Second, the $\varepsilon_2^{interband}(\omega)$ spectra have four peaks (at 1.0, 2.0, 3.1, and 4.2 eV), which are in good agreement with the experimentally observed peak positions shown in Fig. 2(b). Finally, $\varepsilon_1^{interband}(0)$ has a theoretical value of 56, which is in good agreement with the experimental $\varepsilon_1^b(0)$ value of 48.

### *2. Ru film*

Figure 3 shows the optical spectra of Ru. [We were unable to find any previous reports describing the optical studies of Ru.] Drude–Lorentz oscillator fittings for the optical spectra of Ru indicated the presence of four absorption peaks, at 0.6, 1.8, 2.9, and 4.1 eV. As shown in Fig. 3(c), the $\varepsilon_2^{interband}(\omega)$ spectra have a very prominent peak at 2.1 eV and some additional weak peaks around 0.6, 0.9, 3.0, 4.0, and 4.7 eV. The $\varepsilon_2^b(\omega)$ spectra have a very broad peak around 1.6 eV. The high $\gamma_j$ values in Ru might make it difficult to observe separate absorption peaks in $\varepsilon_2^b(\omega)$ spectra, but the overall shapes of both $\varepsilon_1^{interband}(\omega)$ and $\varepsilon_2^{interband}(\omega)$ spectra agree reasonably well with those of $\varepsilon_1^b(\omega)$ and $\varepsilon_2^b(\omega)$, respectively. The $\varepsilon_1^{interband}(0)$ has a theoretical value of 74, which is also in good agreement with the experimental $\varepsilon_1^b(0)$ value of 82.

### 3. Pt film

Figure 4 presents the optical spectra of Pt. According to Weaver,[11] Pt has a strong absorption peak around 0.8 eV, and a peak was reported to appear around 7.4 eV, although this was not clearly seen. Drude–Lorentz oscillator fittings for Pt optical spectra show a very strong peak at 0.8 eV, a broad peak around 7.2 eV, and a few other weak peaks.

The overall shapes of $\hat{\varepsilon}^{interband}(\omega)$ spectra are in reasonable agreement with those of $\hat{\varepsilon}^b(\omega)$, except in the very low frequency region. The $\varepsilon_1^{interband}(0)$ has a theoretical value of 73, which is in reasonable agreement with the experimental $\varepsilon_1^b(0)$ value of 58.

### 4. IrO$_2$ film

Figure 5 presents the measured and calculated optical spectra of IrO$_2$. Goel *et al.*[12] determined the optical constants of IrO$_2$. They observed about five low energy transitions between 1.0 eV and 2.0 eV, and several peak structures at higher levels, between 3.5 eV and 8.0 eV, which they interpreted as *p–d* interband transitions. Compared to earlier works, our $\sigma_1(\omega)$ spectra data, shown in Fig. 5(a), exhibits broad features below the 2.0 eV region and some broad features in the ultraviolet region. Using peak position values from previous literature, we were able to find a reasonably good fit with Eq. (3), as shown by the dashed line in Fig. 5(a).

As shown in Fig. 5(b) and 5(c), $\hat{\varepsilon}^{interband}(\omega)$ agree quite well with $\hat{\varepsilon}^{b}(\omega)$. Their overall shapes are quite similar, except for the small overshoot problem in the predicted $\varepsilon_1^{interband}(\omega)$ curve in Fig. 5(c). Most observed peak features in $\varepsilon_1^{b}(\omega)$ are reproduced in the theoretical prediction. The theoretical $\varepsilon_1^{interband}(0)$ value of 32, is also in very good agreement with the experimental $\varepsilon_1^{b}(0)$ value of. 29.

### C. General tendencies in $\hat{\varepsilon}^{b}(\omega)$

A comparison of experimental $\hat{\varepsilon}^{b}(\omega)$ and theoretical $\hat{\varepsilon}^{interband}(\omega)$ revealed that overall spectra shapes agree fairly well. Experimentally observed peak positions agree particularly well with theoretical predictions, except for Ru. We should note that we observed strong low energy transition peaks at levels lower than 2.0 eV in all samples, in both experiments and calculations. As the number of valence electrons increased, the low energy peaks in Ir, Ru, and Pt exhibited a red shift in position and a narrowing of peak width.

## IV. DISCUSSION

### A. Low energy *d–d* transitions

We examined the density of states (DOS) near the Fermi energy ($E_F$) in an effort to clarify the strong low energy transition peak below 2.0 eV. Figure 6(a) and 6(b) display the calculated total and projected DOS for Ir and $IrO_2$, respectively. [Since Ru and Pt

have DOS plots similar to Ir, we do not discuss them separately.]

Figure 6(a) shows that the most DOS for Ir near the $E_F$ consists of Ir-$5d$ bands. These are located within a rather wide energy region of 9.5 eV. The absorption peaks likely result from interband transitions, from occupied to empty $5d$ bands. A prominent feature appears in the empty band DOS around 0.7 eV above $E_F$. In addition, peaks are evident in the occupied band DOS around $-1.0$, $-1.7$, $-2.8$, $-4.0$, and $-5.6$ eV. Energy differences between these levels and prominent empty band features are about 1.7, 2.4, 3.5, 4.7, and 6.3 eV, respectively. These values are in reasonably good agreement with the experimentally observed peak positions of 1.0, 1.9, 3.1, 4.2, and 6.3 eV.

Figure 6(b) presents the angle-averaged DOS of $IrO_2$, which is composed of three parts: unoccupied Ir-$5d$ $e_g$ bands at 1.0–4.5 eV, partially occupied Ir-$5d$ $t_{2g}$ bands near $E_F$ with a bandwidth of 3.7 eV, and a valence band of hybridized O-$2p$ and Ir-$5d$ states located at $-3.0$ to $-10.0$ eV. The unoccupied Ir-$5d$ $e_g$ band DOS has broad structures from 1.0 eV to 4.5 eV. The Ir-$5d$ $t_{2g}$ band DOS near the $E_F$ exhibits peak structures in the unoccupied band below 1.0 eV, and a prominent peak around $-2.0$ eV in the occupied band. The valence band of hybridized O-$2p$ and Ir-$5d$ states exhibits a broad peak centered at around $-5.0$ eV.

The low energy absorption peaks in $IrO_2$ likely are due to transitions from occupied to unoccupied Ir-$5d$ bands. The prominent low energy absorption peak at 0.4 eV is probably attributable to electronic transitions within the partially occupied Ir-$5d$ $t_{2g}$ band located at $E_F$. Other low energy absorption peaks at 0.9 eV and 1.8 eV are a result of transitions from DOS peaks at $-1.1$ eV and at $-2.0$ eV, respectively, to the unoccupied Ir-$5d$ $t_{2g}$ band. In addition to $d$–$d$ transitions, $p$–$d$ transitions begin at 3.0 eV for $IrO_2$. Absorption peaks at 3.6 eV and 4.3 eV correspond to transitions from O-$2p$ band DOS peaks at −3.5 eV and −4.4 eV, respectively, to the unoccupied Ir-5d $t_{2g}$ band. The absorption peaks at 6.3 eV and 7.2 eV are likely due to transitions from O-2p bands to Ir-5d $e_g$ bands.

As the number of valence electrons increased in transition metals, we observed that 5d state bandwidth narrowed and that the $E_F$ got closer to the DOS peak of unoccupied 5d bands (data not shown). Such phenomena could result in the observed narrowing in peak width and red shift in the prominent low energy peak within the optical spectra, as discussed in Section III C. Our comparison of possible optical transitions in the DOS plot and experimental spectra reveal that the strong transitions below 2.0 eV likely result from d–d transitions.

### B. High $\varepsilon_1^b(0)$ values and their origins

Typical noble metals such as Cu, Ag, and Au have $\varepsilon_1^b(0)$ values of 4.8, 1.5, and 6.9, respectively.[10,19] SrRuO$_3$ is one of the most commonly used electrode materials in FE studies, and has a $\varepsilon_1^b(0)$ value of about 8.2.[4] In contrast, Ir, Ru, Pt, and IrO$_2$ have unexpectedly high $\varepsilon_1^b(0)$ values of 48, 82, 58, and 29, respectively. These values are about one order of magnitude higher than the corresponding values for noble metals.

Equation (4) reveals that $\varepsilon_1^b(0)$ can have a high value when one $n_j$ value is high and its corresponding $\omega_j$ value is low, *i.e.*, at least one strong absorption peak exists in the low energy region. In noble metals, only *s*-electrons contribute to conduction, and other bands are completely filled. The broad nature of the *s*-band does not allow a major contribution to electronic transitions, and the only possible transitions are from filled *d*-bands to other higher bands, which occur at photon energy levels higher than 2.0 eV. Accordingly, noble metals cannot have high $\varepsilon_1^b(0)$ values.

In contrast, this investigation of electrode materials has found strong *d–d* transitions in the energy region lower than 2.0 eV. Drude–Lorentz oscillator fittings of the experimental dielectric functions (summarized in Table I) reveal that oscillators below 2.0 eV make the largest contributions to $\varepsilon_1^b(0)$ values. These oscillators correspond to interband transitions between *d* bands. According to the dipole selection rule, the matrix element of *d–d* transitions in an isolated atom should be zero, so corresponding electric dipole absorption should be forbidden. However, in real solids, hybridization effects and some local structural distortions can make the matrix element nonzero, resulting in *d–d* transitions. Our first-principles calculations reveal quite large matrix element terms for *d–d* transitions in Ir, Ru, Pt, and $IrO_2$. Therefore, strong transitions, accompanied with relatively low transition energy and high DOS values (due to the narrow *d*-band bandwidth), result in unexpectedly high $\varepsilon_1^b(0)$ values for these electrode materials.

### C. Effective mass

The effective mass ($m^*$) of free electrons is another physical quantity required to determine $E_d$ in a FE capacitor. We can determine screening length ($\lambda$) in Eq. (1) from the dielectric constant and effective mass by $\lambda = \lambda_{TF}\sqrt{\dfrac{\varepsilon_e}{m^*}}$, where $\lambda_{TF}$ is the Thomas–Fermi screening length. Using our calculated density of states at $E_F$ and experimental specific heat values,[20–22] we roughly estimated effective mass; Table II presents the results. The $m^*/m$ values of these electrode materials range from 1.4 to 1.7, which are fairly close to 1.

### D. Implications for ferroelectric capacitor applications

An electrode material's high $\varepsilon_1^b(0)$ value has some important implications for actual applications in capacitor-type FE devices. As a device's size is reduced, its FE film thickness also decreases. According to Eq. (1), depolarization field $E_d$ grows with a decrease in film thickness. A large depolarization field could result in critical thickness, when film ferroelectricity vanishes since FE double-well potential disappears due to $E_d$.[3] In a previous study, we found that a large $E_d$ could also result in a rapid decrease in polarization by creating domains with opposite polarization.[5] This relaxation in polarization could create another fundamental limit for capacitor-type FE devices.[4] Since $E_d$ results from an incomplete screening of polarization charge at the metal/FE interface, its occurrence is inevitable in capacitor-type FE devices. However, by choosing an electrode material with high $\varepsilon_1^b(0)$ values, researchers could reduce $E_d$ and create better FE devices.

Until now, selection of electrode materials for FE capacitors has been based on improving device performance and resolving reliability issues. Noble metals produce known problems such as adhesion. Furthermore, when a Pt electrode is used, FE materials such as PZT exhibit a fatigue problem.[23] Because of these flaws, other materials such as Ir, Ru, and $IrO_2$ are generally used in electrodes. Our research indicates that these electrode materials produce higher $\varepsilon_1^b(0)$ values than noble metals. Therefore, these transition metals or oxides are a better choice for electrode materials when the thickness of a FE device is approaching its fundamental limit.

## V. SUMMARY

We obtained the dielectric constants of bound charges for transition metals (Pt, Ir, and, Ru) and transition metal dioxide ($IrO_2$), which are typically used as metal electrodes in electronic devices. From the dielectric function spectra attained from optical spectroscopy, free carrier contributions were separated to give only the contribution from bound electrons. These contributions from bound electrons to the *dc* dielectric constant agree with predictions based on first-principles calculations. In particular, we observed pronounced low energy *d–d* transitions in both experimental and theoretical optical spectra. These low energy transitions in the transition metals and $IrO_2$ are responsible for the high *dc* dielectric constant values of the bound charge. We also examined the advantageous implications of high dielectric constant values in actual device applications.

## ACKNOWLEDGMENTS

We thank D. J. Kim, Y. D. Park, and K. S. Suh for their valuable input. This study was financially supported by Creative Research Initiatives (Functionally Integrated Oxide Heterostructures) from the Ministry of Science and Technology (MOST)/Korean Science and Engineering Foundation (KOSEF). Experiments conducted at Pohang Light Source were supported in part by MOST and Pohang University of Science and Technology (POSTECH). W.S.C. acknowledges support from the Seoul City. Calculations were supported by the Korea Institute of Science and Technology Information (KISTI) through the $7^{th}$ Strategic Supercomputing Support Program.

**FIGURES & TABLES**

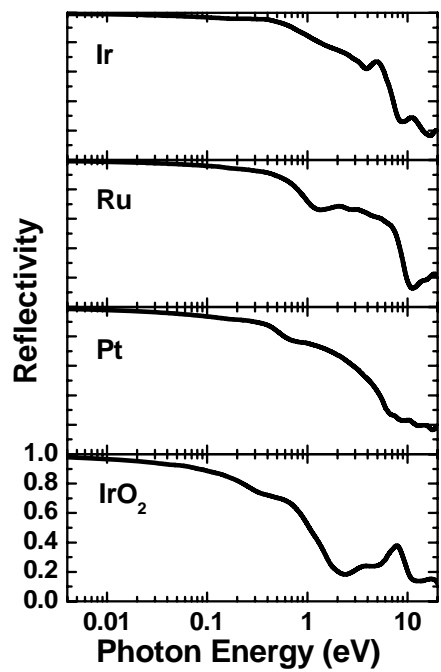

FIG. 1. Reflectance spectra of Ir, Ru, Pt, and $IrO_2$.

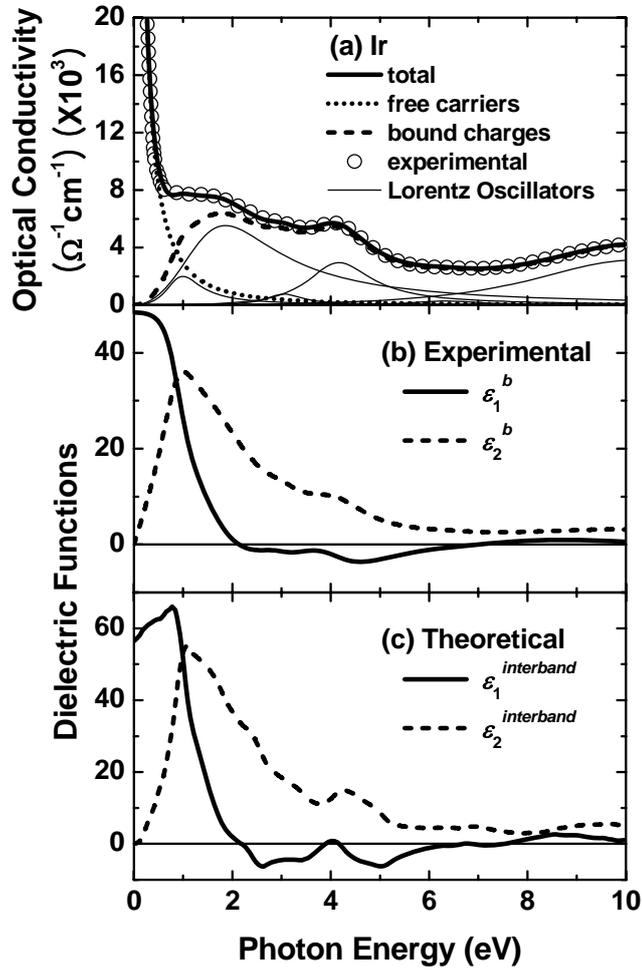

FIG. 2. Optical spectra of Ir. (a) Optical conductivity spectra: experimental results (represented by open circles), total fit of optical conductivity (thick solid line), contribution from free carriers (dotted line), and contribution from bound charges (dashed line). Thin solid lines represent the contributions of each Lorentz oscillator. (b) Real part (represented by the solid line) and imaginary part (dashed line) of the experimental fit results for $\varepsilon^b(\omega)$. (c) Real part (represented by the solid line) and imaginary part (dashed line) of the theoretical results for $\varepsilon^{interband}(\omega)$.

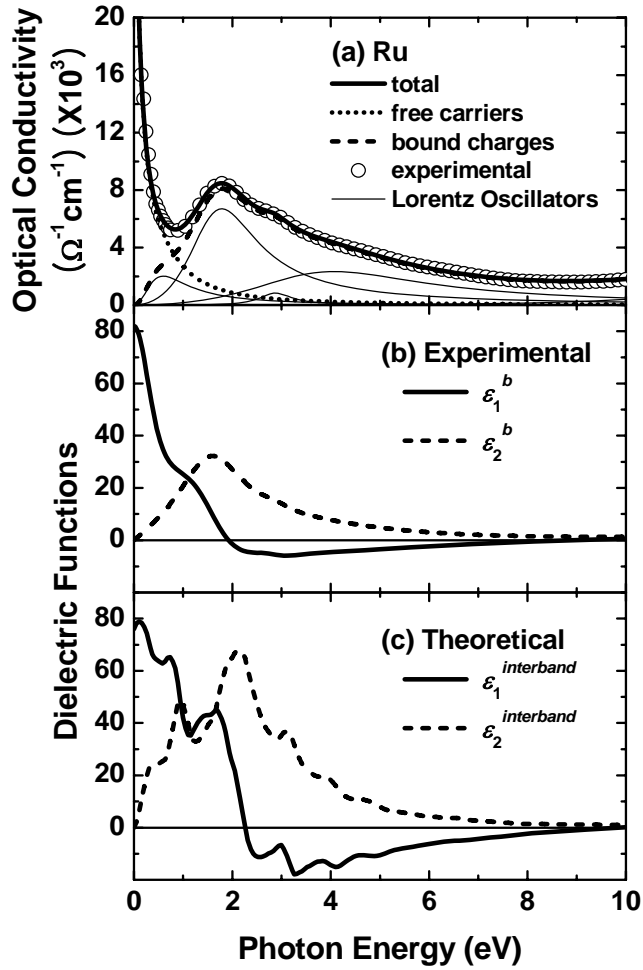

FIG. 3. Optical spectra of Ru. (a) Optical conductivity spectra: experimental result (represented by open circles), total fit of optical conductivity (thick solid line), contribution from free carriers (dotted line), and contribution from bound charges (dashed line). Thin solid lines represent contributions of each Lorentz oscillator. (b) Real part (represented by the solid line) and imaginary part (dashed line) of the experimental fit results for $\varepsilon^{b}(\omega)$. (c) Real part (represented by the solid line) and imaginary part (dashed line) of the theoretical results for $\varepsilon^{interband}(\omega)$.

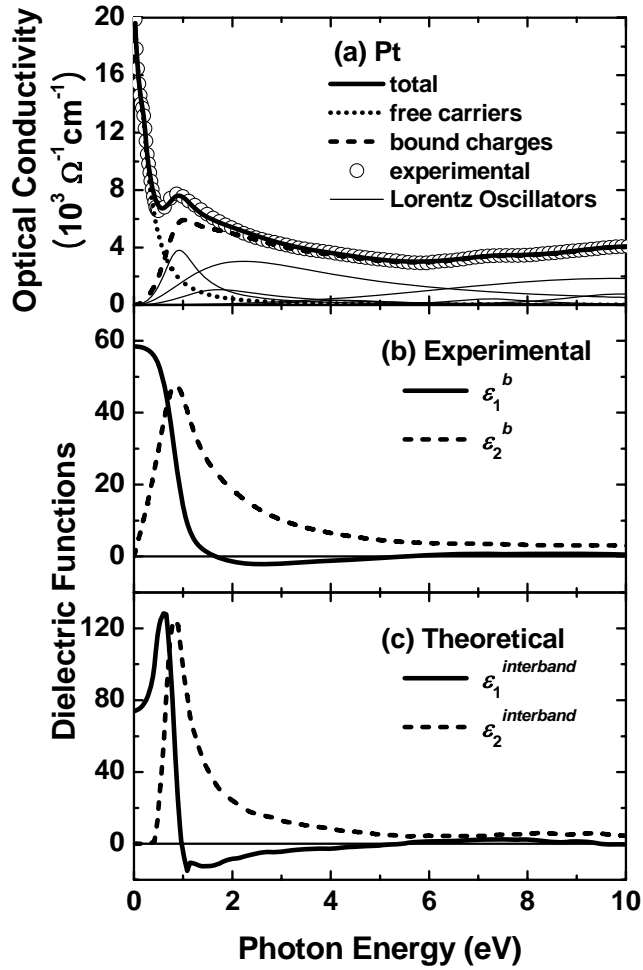

FIG. 4. Optical spectra of Pt. (a) Optical conductivity spectra: experimental result (represented by open circles), total fit of optical conductivity (thick solid line), contribution from free carriers (dotted line), and contribution from bound charges (dashed line). Thin solid lines represent contributions of each Lorentz oscillator. (b) Real part (represented by the solid line) and imaginary part (dashed line) of the experimental fit results for $\varepsilon^b(\omega)$. (c) Real part (represented by the solid line) and imaginary part (dashed line) of the theoretical results for $\varepsilon^{interband}(\omega)$.

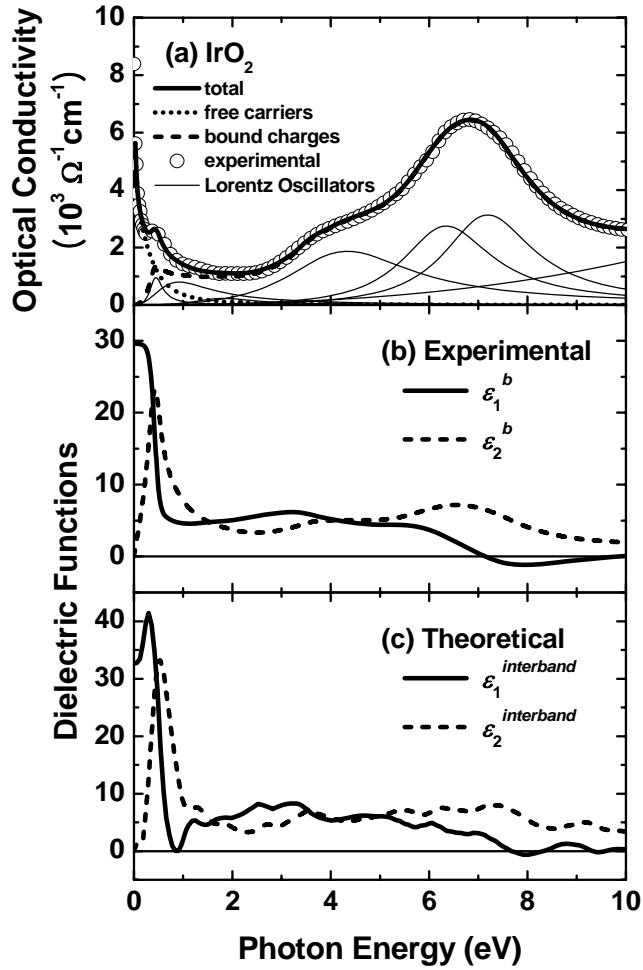

FIG. 5. Optical spectra of IrO$_2$. (a) Optical conductivity spectra: experimental result (represented by open circles), total fit of optical conductivity (thick solid line), contribution from free carriers (dotted line), and contribution from bound charges (dashed line). Thin solid lines represent contributions of each Lorentz oscillator. (b) Real part (represented by the solid line) and imaginary part (dashed line) of the experimental fit results for $\varepsilon^b(\omega)$. (c) Real part (represented by the solid line) and imaginary part (dashed line) of the theoretical results for $\varepsilon^{interband}(\omega)$.

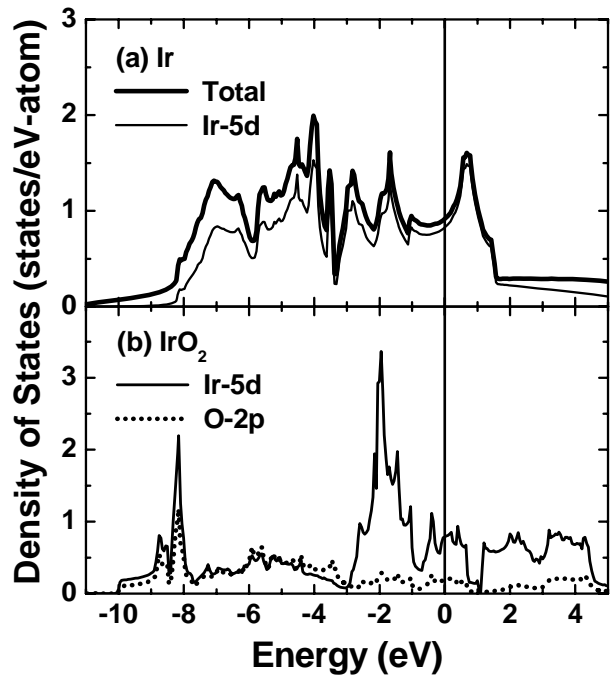

FIG. 6. Density of states (DOS) of (a) Ir and (b) $IrO_2$. The thick solid line in (a) represents the total DOS. The thin solid lines in (a) and (b), and dashed lines in (b) represent the projected DOS of Ir-$5d$ and O-$2p$ states, respectively. Fermi energy was set at zero and is indicated by the vertical solid line.

TABLE I. Parameter values of Lorentz oscillators for Ir, Ru, Pt, and IrO$_2$. The individual contribution each Lorentz oscillator makes to $\varepsilon_1^b(0)$ is presented as $\Delta\varepsilon_1^b(0)$, where $\varepsilon_\infty$ corresponds to the summation of contributions from transitions greater than 10.0 eV.

| Material | $\omega_j$ (eV) | $\omega_{pj}$ (eV) | $\gamma_j$ (eV) | $\Delta\varepsilon_1^b(0)$ |
|---|---|---|---|---|
| Ir | 1.0 | 3.6 | 0.9 | 13.1 |
|  | 1.9 | 10.1 | 2.5 | 29.5 |
|  | 3.1 | 2.3 | 0.9 | 0.6 |
|  | 4.2 | 5.6 | 1.6 | 2.0 |
|  | 6.3 | 1.9 | 1.6 | 0.1 |
|  |  |  |  | $\varepsilon_\infty = 3.1$ |
| Ru | 0.6 | 4.1 | 1.1 | 47.3 |
|  | 1.8 | 9.3 | 1.7 | 27.1 |
|  | 2.9 | 2.2 | 0.8 | 0.6 |
|  | 4.1 | 9.1 | 4.5 | 5.0 |
|  |  |  |  | $\varepsilon_\infty = 2.0$ |
| Pt | 0.9 | 5.0 | 0.9 | 29.2 |
|  | 1.7 | 4.0 | 2.2 | 5.4 |
|  | 2.2 | 10.0 | 4.3 | 19.8 |
|  | 4.2 | 2.5 | 2.5 | 0.4 |
|  | 7.2 | 2.6 | 2.1 | 0.1 |
|  | 9.8 | 12.4 | 11.2 | 1.6 |
|  | 9.9 | 4.1 | 2.9 | 0.2 |
|  |  |  |  | $\varepsilon_\infty = 1.9$ |
| IrO$_2$ | 0.4 | 1.4 | 0.3 | 10.2 |
|  | 0.9 | 3.0 | 1.5 | 11.8 |
|  | 1.8 | 1.6 | 1.5 | 0.8 |
|  | 3.6 | 0.9 | 0.6 | 0.1 |
|  | 4.3 | 6.6 | 3.1 | 2.3 |
|  | 6.3 | 6.8 | 2.2 | 1.1 |
|  | 7.2 | 7.1 | 2.2 | 1.0 |
|  |  |  |  | $\varepsilon_\infty = 2.4$ |

TABLE II. Experimental ($\varepsilon_1^b(0)$) and theoretical ($\varepsilon_1^{interband}(0)$) dielectric constants and effective mass ratios ($m^*/m$) calculated for Ir, Ru, Pt, and IrO$_2$.

|  | Ir | Ru | Pt | IrO$_2$ |
|---|---|---|---|---|
| $\varepsilon_1^b(0)$ | 48 ± 10 | 82 ± 10 | 58 ± 10 | 29 ± 5 |
| $\varepsilon_1^{interband}(0)$ | 56 | 74 | 73 | 32 |
| $m^*/m$ | 1.4–1.5 | 1.7 | 1.6–1.7 | 1.62 |